\documentclass[10pt,conference]{IEEEtran}
\IEEEoverridecommandlockouts
\usepackage{hyperref}
\usepackage{cite}
\usepackage{amsmath,amssymb,amsfonts}
\usepackage{graphicx}
\usepackage{dblfloatfix}
\usepackage{textcomp}
\usepackage{xcolor}
\usepackage{url}
\usepackage{booktabs}
\usepackage{tcolorbox}
\usepackage{multirow}
\usepackage{ulem}
\usepackage{balance}
\usepackage{listings}
\usepackage{enumitem}
\usepackage{comment}
\usepackage{siunitx}
\usepackage{amssymb}
\usepackage{rotating}
\usepackage{longtable}
\usepackage{tabularx,booktabs}
\usepackage{lipsum}
\usepackage{hyperref}
\usepackage{microtype} 
\usepackage{xspace}
\usepackage{algorithm}
\usepackage{algpseudocode}
\usepackage{stmaryrd}
\usepackage{soul}
\usepackage{color, colortbl}
\usepackage{subfigure}
\usepackage{caption}
\usepackage{epigraph} 
\usepackage{framed}
\usepackage{here}
\definecolor{coolblack}{rgb}{0.0, 0.18, 0.39}
\usepackage{xurl}
\definecolor{awesome}{rgb}{0.0, 0.2, 0.6}
\usepackage{datenumber}

\newcounter{dateone}\newcounter{datetwo}%
\newcommand{\daydifftoday}[3]{%
\setmydatenumber{dateone}{\the\year}{\the\month}{\the\day}%
\setmydatenumber{datetwo}{#1}{#2}{#3}%
\addtocounter{datetwo}{-\thedateone}%
\thedatetwo
}

\usepackage{float}

\newcommand{\RqOne}{\textbf{RQ1:} \emph{To what extent is a deprecated Log4j version still being used in software projects?}}
\newcommand{\RqTwo}{\textbf{RQ2:} \emph{ Do newcomer projects tend to adopt the newer version or the deprecated version?}}
\newcommand{\RqThree}{\textbf{RQ3:} \emph{Are software projects with more releases more likely to adopt the new version over the deprecated version?}}

\def\BibTeX{{\rm B\kern-.05em{\sc i\kern-.025em b}\kern-.08em
    T\kern-.1667em\lower.7ex\hbox{E}\kern-.125emX}}
    
\begin{document}

\title{Do Developers Depend on Deprecated Library Versions? A Mining Study of Log4j}

\author{\IEEEauthorblockN{Haruhiko Yoshioka $^1$, Sila Lertbanjongngam$^1$, Masayuki Inaba$^1$, Youmei Fan$^1$, \\ Takashi Nakano$^1$, Kazumasa Shimari$^1$, Raula Gaikovina Kula$^2$, Kenichi Matsumoto$^1$}
	\IEEEauthorblockA{
		$^1$\textit{Graduate School of Science and Technology, Nara Institute of Science and Technology}\\
        $^2$\textit{Graduate School of Information Science and Technology, Osaka University}\\
		\{yoshioka.haruhiko.yi4, lertbanjongngam.sila.lo9, inaba.masayuki.iq4\}@naist.ac.jp, \\\{fan.youmei.fs2, nakano.takashi.nr1, k.shimari, matumoto\}@is.naist.jp,\\ raula-k@ist.osaka-u.ac.jp}
}

\maketitle

\begin{abstract}
Log4j has become a widely adopted logging library for Java programs due to its long history and high reliability.
Its widespread use is notable not only because of its maturity but also due to the complexity and depth of its features, which have made it an essential tool for many developers.
However, Log4j 1.x, which reached its end of support (deprecated), poses significant security risks and has numerous deprecated features that can be exploited by attackers. Despite this, some clients may still rely on this library.
We aim to understand whether clients are still using Log4j 1.x despite its official support ending. 
We utilized the Mining Software Repositories 2025 challenge dataset, which provides a large and representative sample of open-source software projects.
We analyzed over 10,000 log entries from the Mining Software Repositories 2025 challenge dataset using the Goblin framework to identify trends in usage rates for both Log4j 1.x and Log4j-core 2.x. Specifically, our study addressed two key issues:
(1) We examined the usage rates and trends for these two libraries, highlighting any notable differences or patterns in their adoption.
(2) We demonstrate that projects initiated after a deprecated library has reached the end of its support lifecycle can still maintain significant popularity.
These findings highlight how deprecated are still popular, with the next step being to understand the reasoning behind these adoptions.

\end{abstract}

\begin{IEEEkeywords}
Log4j, Security Vulnerabilities, Library Migration
\end{IEEEkeywords}

\section{Introduction}

Like many other libraries, Log4j, a popular library, also faces deprecation. Log4j 1.x has known security vulnerabilities and is no longer supported. As a result, it's strongly recommended to upgrade to Log4j-Core v2 to reduce security risks and maintain the stability of software systems.
A key vulnerability, CVE-2019-17571 in Log4j 1.x, has raised serious security concerns and led many projects to migrate to newer versions~\cite{He2021}.
Notwithstanding the security risks associated with Log4j 1.x, many projects may still be using this library due to various factors such as technical debt, compatibility issues, or lack of awareness. This study aims to shed light on the reasons behind the stagnation in migration efforts by analyzing the adoption status of Log4j 1.x and Log4j-Core v2.

Security risks stemming from the Log4j v1 vulnerability and migration to mitigate its impact have been the focus of much attention. Previous studies have found the following findings:
Library dependency updates: investigating how quickly developers update dependencies after security advisories~\cite{Raula2018}.
Delays in fixing vulnerabilities in the npm ecosystem: analyzing delays in fix releases, adoption by direct dependents, and propagation to indirect dependents~\cite{Bodin2021}.
Additionally, recent research highlights broader trends in Java logging practices, such as shifts from ad-hoc libraries like Log4j v1 to abstraction libraries (e.g., slf4j) and unified solutions like Log4j-Core v2~\cite{He2021}. These studies suggest that migrations are not only driven by security concerns but also by factors such as usability, performance, and maintenance challenges. 
Notably, developers often cite the flexibility of newer logging libraries and their ability to simplify integration with modern software ecosystems as key reasons for migration~\cite{Kabinna2016AnEO}.

\begin{figure*}[t]
	\centering
	\includegraphics[width=\textwidth]{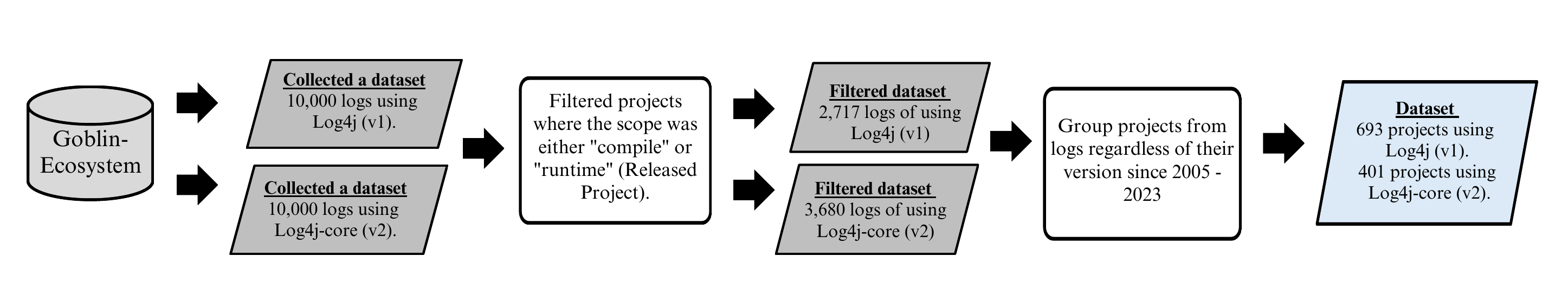}
	\caption{Overview of the data collection and preparation for the study.}
	\label{fig:data-preparation-overviwe}
\end{figure*}

Another effect of managing libraries is the risk of deprecation of library versions.
As a consequence, developers frequently struggle with managing deprecated APIs, an issue explored in various contexts including Python libraries~\cite{Wang2020FSE} and Java ecosystems~\cite{Zhou2016FSE, Sawant2018EMSE}. 
Other studies show that guidance provided for such deprecated APIs often lacks clarity, as evidenced by the inconsistent use of replacement messages~\cite{Brito2018JSS}, thereby contributing to a state of technical lag in package dependency networks~\cite{Decan2018ICSME}.
There has also been work that highlight the difficuilty of maintaining secure, up-to-date software, especially when vulnerabilities are propagated through dependent libraries~\cite{Liu2020ICSE,Raula2018,Lauinger2017NDSS}. 

For this study, our key goal for the 2025 mining challenge is to explore whether or not such libraries that are at the end of life are deprecated.
In particular, our focus is on examining the characteristics of projects that have not made the transition from Log4j 1.x to Log4j-Core v2. We will investigate factors such as project size, complexity, industry, and technology stack to identify potential barriers to migration and understand the challenges developers face when migrating critical libraries.
By gaining insights into these factors, we hope to contribute to a better understanding of the migration process and provide guidance for project leaders and developers on how to prioritize security and migration efforts in their software development projects.
This study will mine and extract migration patterns from Log4j v1 to Log4j-Core v2. \\

This study addresses the following three research questions:

\noindent
-- \RqOne

By answering this research question, we aim to investigate the extent to which a deprecated library like Log4j version 1 is still being used, and to identify the characteristics of software projects that are most likely to continue using it. This will provide valuable insights into the library migration patterns and help us better understand the factors that influence the adoption and use of technology in modern software development.
This study aims to quantify the usage of Log4j v1.

\noindent
-- \RqTwo

This research question aims to analyze the extend by which projects that were created after the end of life was announced (newcomer projects) will either adopt the new version over the deprecated version.

\noindent
-- \RqThree

The final research question investigates whether or not projects that tend to generate more releases (release frequency).

The replication package is available at GitHub Repository\footnote{\url{https://github.com/polarbearpppp/Log4j-Msr2025}}.

\section{Study Design}
In this section, we present the data preparation and approach to answer the two research questions.

\subsection{Data Preparation}

Fig. \ref{fig:data-preparation-overviwe} shows the flow of data preparation.
The cylinder denotes a system, the diamond represents a dataset, and the rectangle indicates data processing.
Initially, 10,000 logs utilizing Log4j v1 were extracted from the Goblin ecosystem, and the same procedure was applied to logs utilizing Log4j-Core v2.
Subsequently, projects corresponding to released applications were identified. 

We examined a subset of 10,000 logs for this investigation. In order to guarantee that the dataset is representative and able to successfully capture patterns, trends, and anomalies, this sample size was selected. It is consistent with log analysis guidelines, which typically suggest a sample size of 10,000–50,000 logs for preliminary analyses. 
The sample size of 10,000 logs is sufficient to minimize variability and sampling error, ensuring robust findings without unnecessary oversampling.
An artifact was considered part of a released project if its scope was designated as either "compile" or "runtime."
After this filtering process, the dataset was reduced to 2,717 logs for Log4j v1 and 3,680 logs for Log4j-Core v2.
Finally, by consolidating logs from 2005 to 2023 and grouping them by project, 693 unique projects using Log4j v1 and 401 unique projects using Log4j-Core v2 were identified.

\subsection {\RqOne}

\textbf{Trend Analysis}. 
To answer RQ1, we identify the common characteristics of projects that still rely on Log4j v1. First, we gathered and analyzed the artifacts associated with the projects and systems actually in operation. This involved examining the project's source code repository, build scripts, and deployment configurations to gather information about the libraries and frameworks used.

To determine whether an artifact is part of a release application, we examined the "Scope" property associated with each artifact. Specifically, we checked if the scope was set to either "compile" or "runtime", which indicated that the artifact was intended for inclusion in the final release application. Conversely, if the scope was set to "test", the artifact was excluded from the release application. By applying this filtering process, we could determine whether an artifact was part of a release application and analyze its characteristics more closely.
By accurately classifying artifacts based on their scope properties, we were able to identify a subset of projects that still relied on Log4j v1. Our analysis focused on the scope-related metadata associated with each artifact, which provided valuable insights into the projects' maintenance practices and deployment configurations. 
To show our research, we use a scatter plot to understand the trend of usage (usage trend) over time.
In detail, we can visualize the distribution of the latest release dates of projects using Log4j v1 over time.
The method involves extracting Artifact nodes and their associated Release nodes from the Goblin dataset.

\subsection{\RqTwo}



\textbf{Number of Newcomer Vs Continuing Projects using Log4j}.
The objective is to determine the proportion of projects using Log4j v1 that are newcomers versus those continuing from prior years.
It also aims to analyze trends in project adoption and the persistence of legacy projects despite the end-of-life (EOL) announcement in 2015.
The method involves examining Artifact and Release nodes from the Goblin dataset~\cite{Damien2025} to identify newcomer and continuing projects based on their release history.
Newcomers are defined as projects adopting Log4j v1 for the first time in a given year, while continuing projects are those carried over from previous years.
The results will be visualized using a bar chart to illustrate adoption trends over time.

\subsection{\RqThree}
\textbf{Frequency of Released Project Activties}.
The objective of this study is to analyze the frequency of release updates among projects using Log4j v1, as frequent releases serve as an indicator of active project maintenance. The methodology involves calculating the intervals between releases for each project based on the timestamp attribute, grouping projects by release frequency to assess their activity levels. Additionally, the study aims to determine the extent of migration from Log4j v1 to Log4j-core v2. This is achieved by matching projects using Log4j v1 with corresponding projects using Log4j-core v2, identifying migration occurrences when matches are found. The results are visualized through scatter plots and heatmaps, illustrating the frequency of release activities and the migration amount from Log4j v1 to Log4j-core v2.

\section{Results}
We now present the results of the study.

\begin{figure}[h]
	\centering
	\includegraphics[width=.45\textwidth]{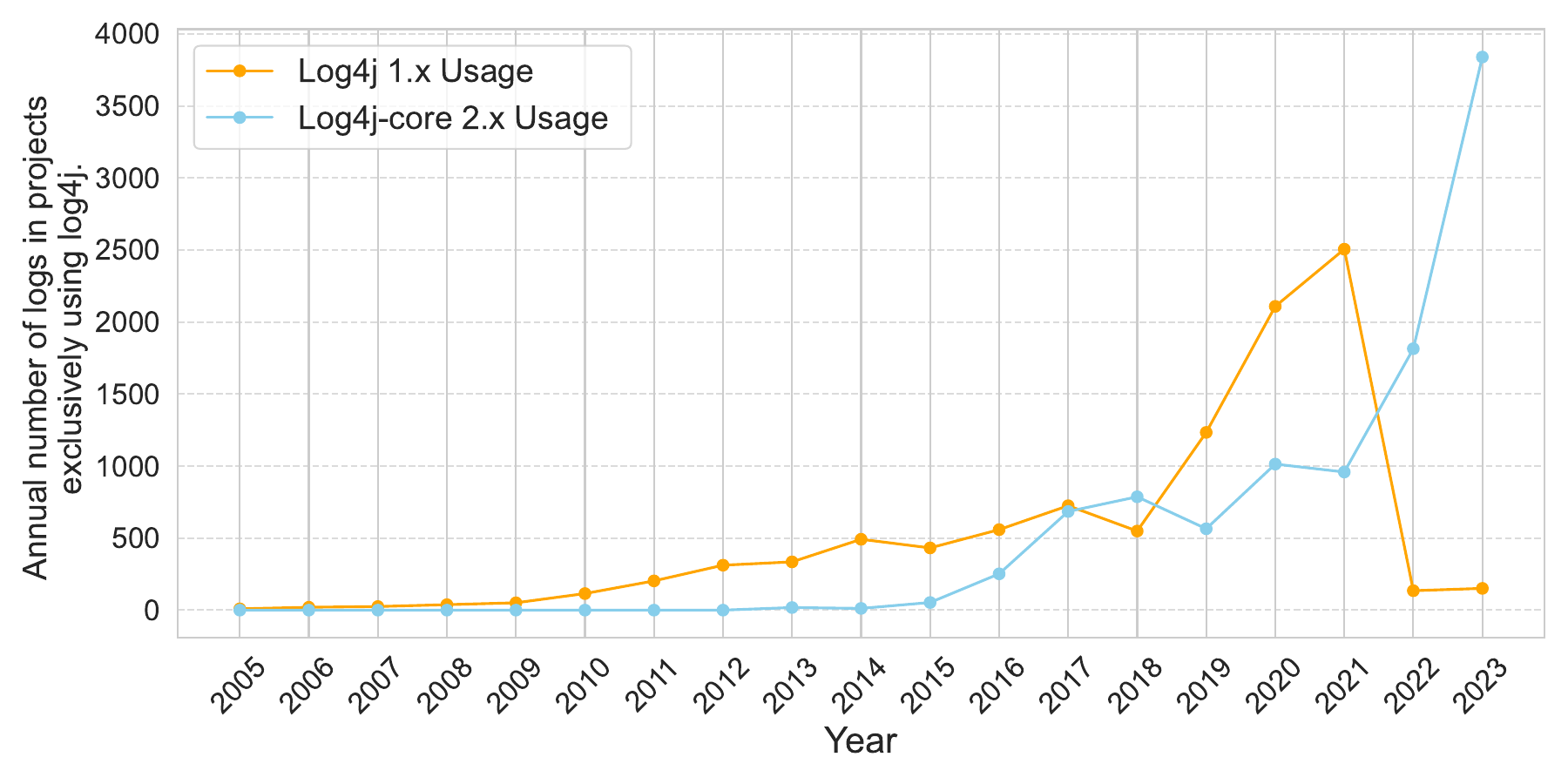}
	\caption{Usage trend of Log4j artifacts over the years, including unreleased projects.}
	\label{fig:log-usage}
\end{figure}

\begin{figure}[h]
	\centering
	\includegraphics[width=.45\textwidth]{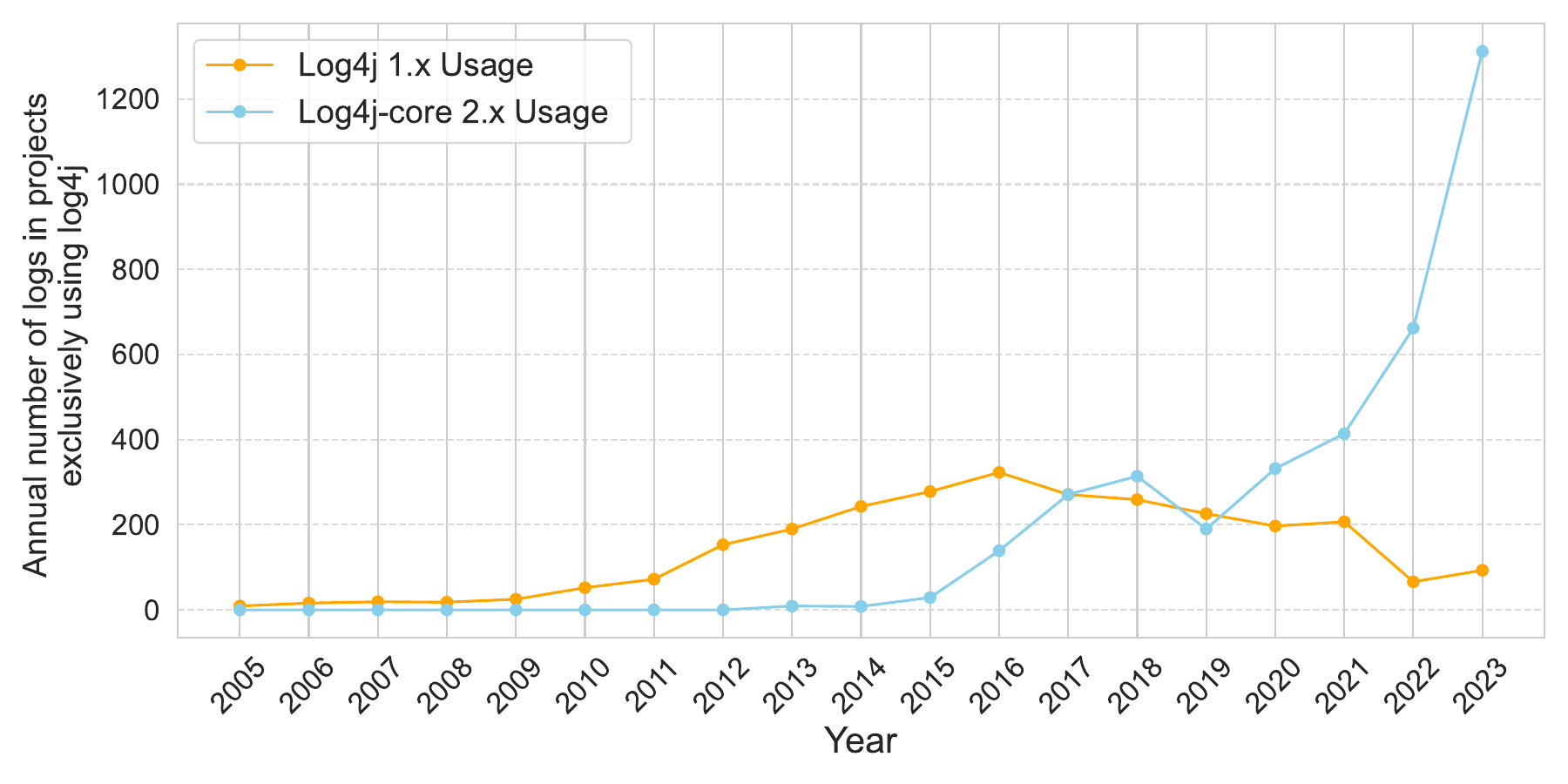}
	\caption{Usage trend of Log4j artifacts over the years, filtered to include only released projects.}
	\label{fig:log-usage_release}
\end{figure}

\begin{figure}[t]
	\centering
	\includegraphics[width=.45\textwidth]{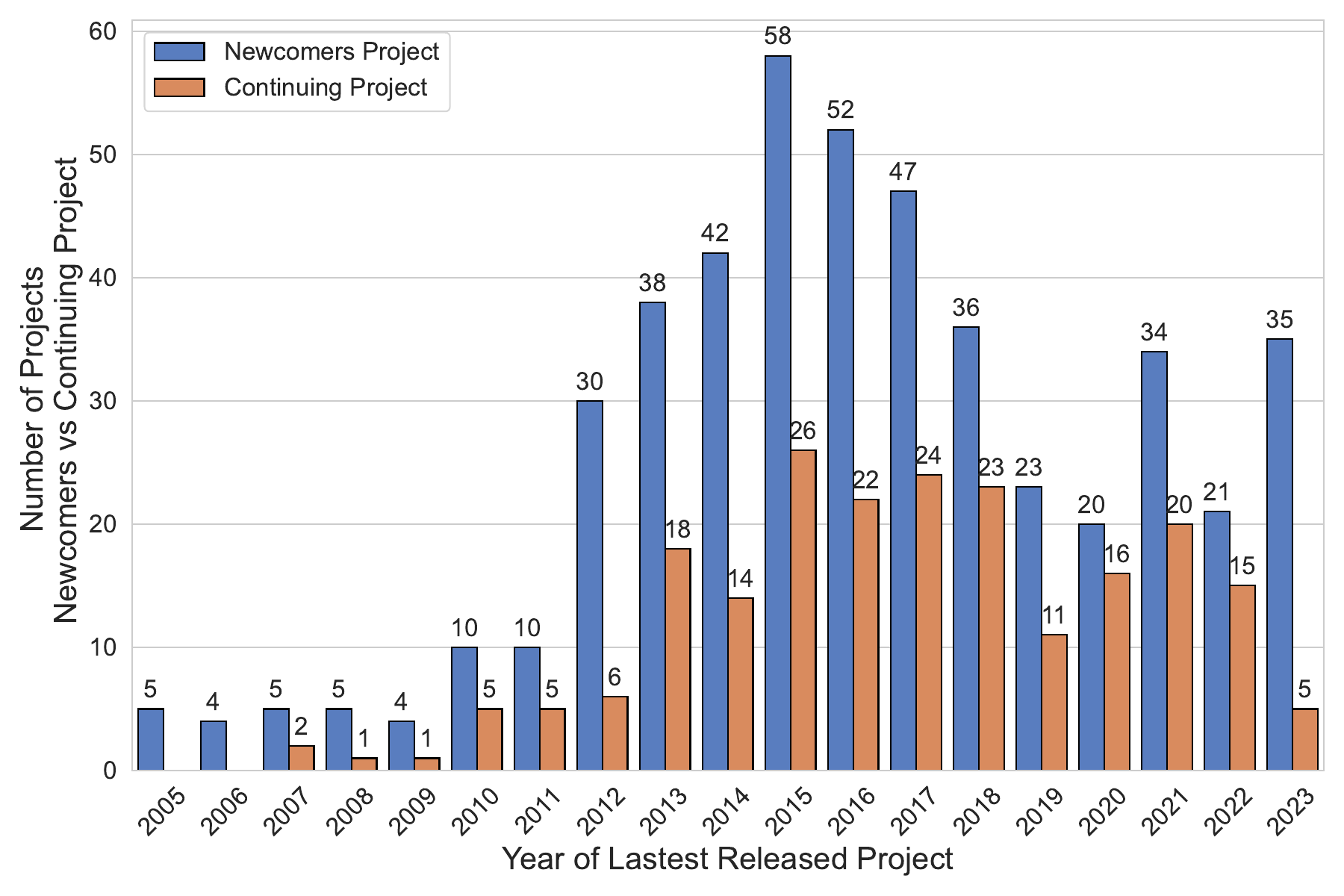}
	\caption{Proportion of newcomers and continuing users of Log4j v1 in latest released projects by year}
	\label{fig:Newcomers_vs_Continuing}
\end{figure}

\begin{figure}[t]
	\centering
	\includegraphics[width=.45\textwidth]{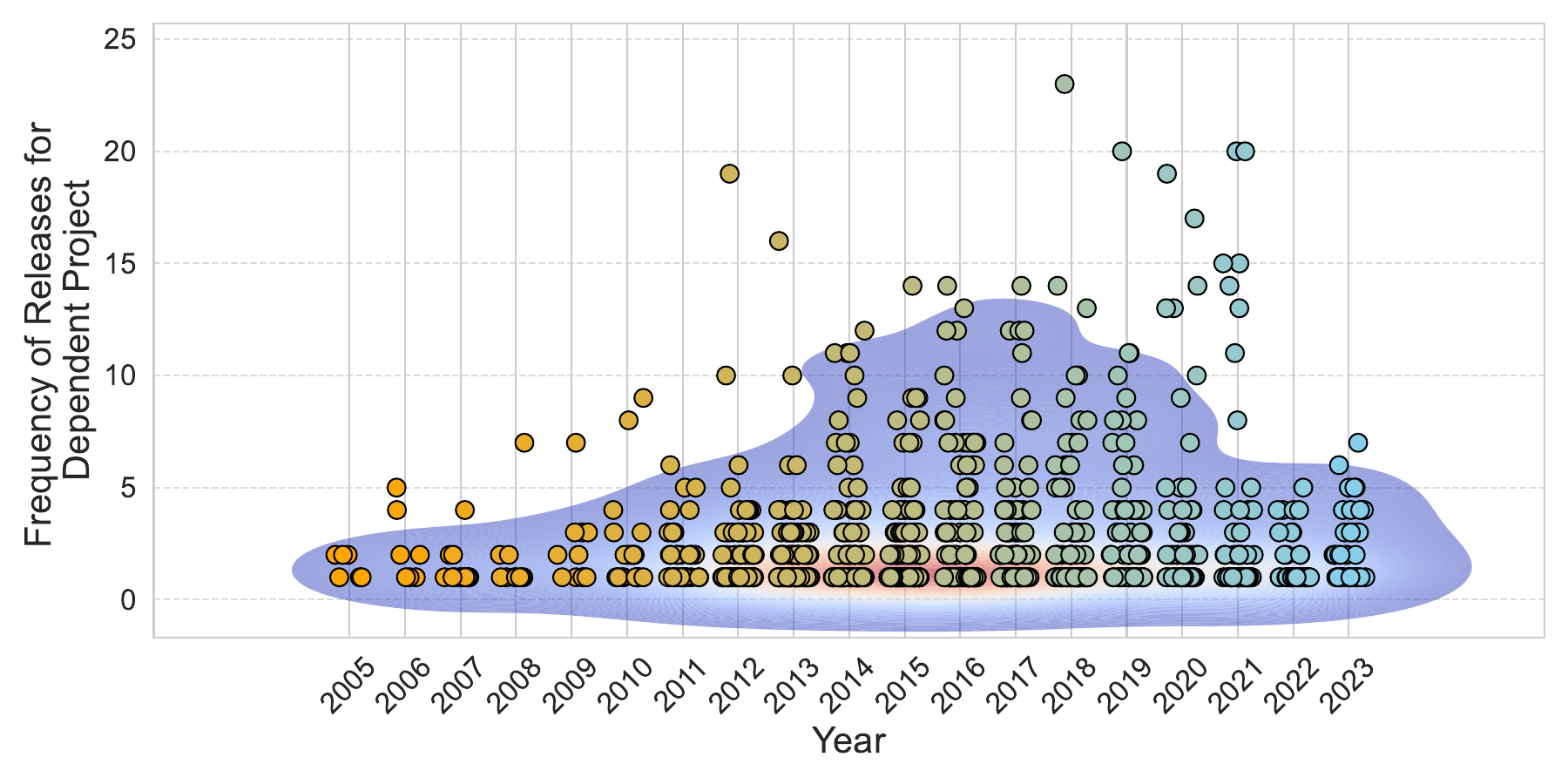}
	\caption{Scatter plot of project releases frequencies with a heatmap.}
	\label{fig:Released_Project_activites}
\end{figure}

\begin{figure*}[h]
	\centering
	\includegraphics[width=.7\textwidth]{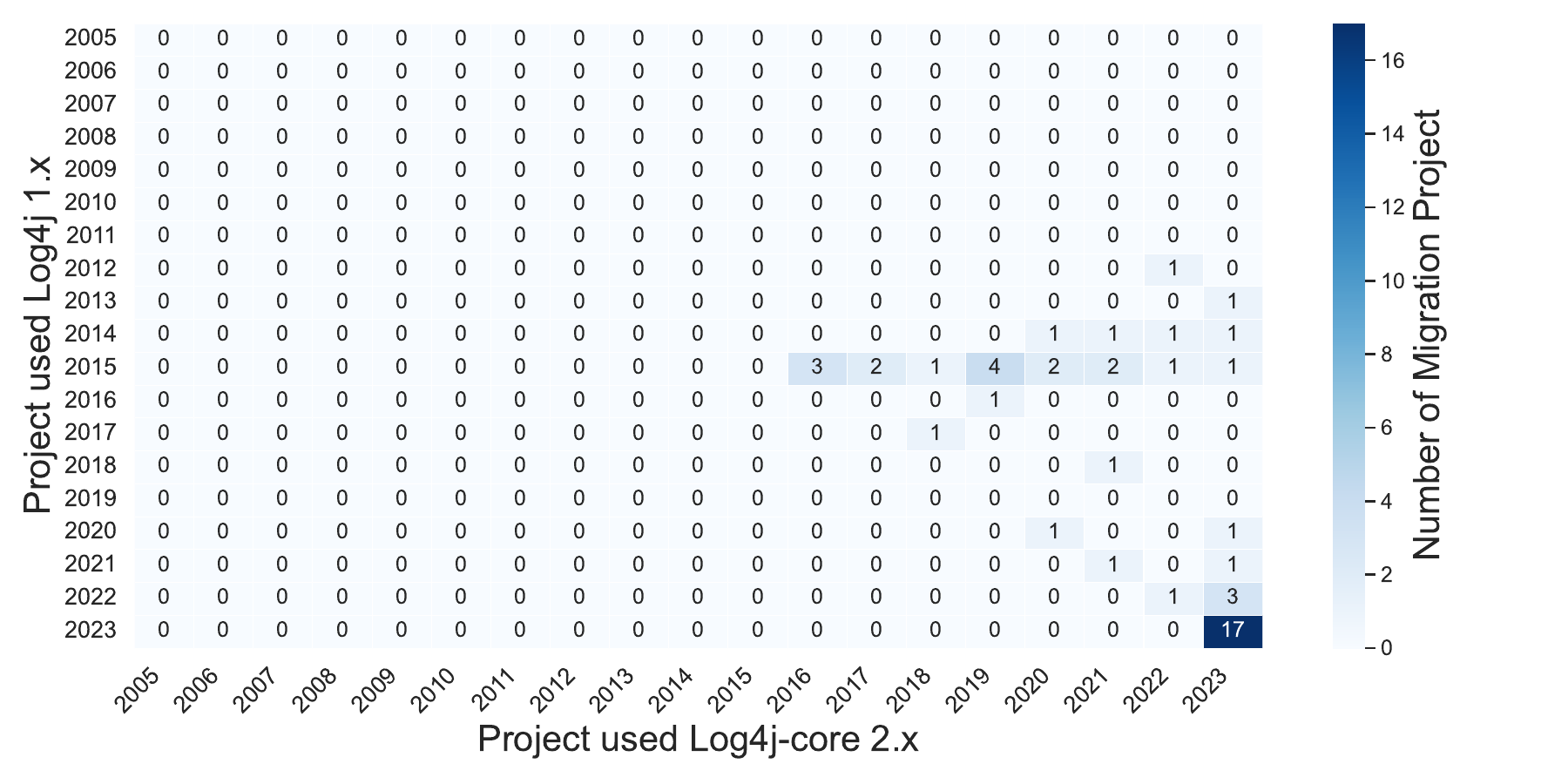}
	\caption{Heatmap showing the number of migrations from Log4j 1.x to Log4j-core 2.x, highlighting the amout of migration activities across projects.}
	\label{fig:Migration_number}
\end{figure*}

\subsection{\RqOne}

From 10,000 logs collected between 2005 and 2023 for both Log4j v1 and Log4j-core v2, the usage trend graph shown in Fig.~\ref{fig:log-usage} illustrates the distribution of counts over the years.

In addition, analysis of the logs reveals the following:
From 10,000 logs of each Log4j 1.x and Log4j-core, the number of artifacts using Log4j v1 and released is approximately \textbf{2,717}.
For Log4j-core v2, the corresponding count is around \textbf{3,680}.
The trend graph in Fig.~\ref{fig:log-usage_release}, illustrates the usage trends of released projects utilizing Java Log4j artifacts over the years. It shows that following the EOL announcement in 2015, the adoption of Log4j-core 2.x increased steadily, while the usage of Log4j 1.x began to decline. However, the figure also highlights that some projects continue to rely on Log4j 1.x despite its deprecated status, indicating ongoing use of the outdated version.

\begin{tcolorbox}[colback=gray!5,colframe=awesome,title= RQ1 Summary]
Deprecated artifacts are still trendy.
Our results indicate that although Log4j-Core v2 has been widely adopted, Log4j v1 still remains popular among released artifacts.
\end{tcolorbox}

\subsection{\RqTwo}




Fig.~\ref{fig:Newcomers_vs_Continuing}  reveals across all 693 projects, an average of (\SI{73.31}{\percent}) were newcomers adopting Log4j for the first time, while the remaining (\SI{26.69}{\percent}) were continuations from prior years. 
Post-2015 data reveals that 404 projects were released after the EOL announcement, of which (\SI{66.19}{\percent}) were new adopters, while (\SI{33.81}{\percent}) continued relying on the older Log4j version.
This persistence highlights that, despite the end-of-life status and associated risks, a significant proportion of projects maintained their reliance on the deprecated Log4j version.

\begin{tcolorbox}[colback=gray!5,colframe=awesome,title= RQ2 Summary]
Interestingly new projects (newcomers) are still adopting the deprecated project, even after the annoucement.
\end{tcolorbox}

\subsection{\RqThree}

Fig.~\ref{fig:Released_Project_activites} presents an investigation into the frequency of activities within released projects over the years, illustrating trends in project activity levels. 
We observe variations in project activity frequency, with peaks and troughs corresponding to specific years. This trend provides insights into the maintenance and release patterns of projects, potentially reflecting periods of increased development focus or stagnation. The data can help identify active versus legacy projects and assess the overall lifecycle dynamics within the examined dataset and we also investigated the migration behavior from Log4j 1.x to Log4j-core 2.x, as depicted in Fig.~\ref{fig:Migration_number}, shows that out of 693 projects analyzed between 2005 and 2023, only 50 projects (\SI{7.21}{\percent}) migrated to the newer version. Notably, more than \SI{88}{\percent} of these migrations occurred after the official end-of-life (EOL) announcement for Log4j 1.x in 2015. 

The analysis of project release frequencies (Fig.~\ref{fig:Released_Project_activites}) and migration patterns from Log4j 1.x to Log4j-core 2.x (Fig.~\ref{fig:Migration_number}) reveals that Despite notable activity levels after the EOL announcement, only 44 out of 404 projects (\SI{10.89}{\percent}) transitioned to Log4j-core 2.x. This indicates that, while projects remain active, the majority have not migrated to the updated version. 

\begin{tcolorbox}[colback=gray!5,colframe=awesome,title= RQ3 Summary]
Despite the end-of-life (EOL) announcement for Log4j 1.x in 2015, \SI{33.81}{\percent} of projects continued using the outdated version post-EOL, while only (\SI{10.89}{\percent}) migrated to Log4j-core 2.x. 

\end{tcolorbox}

\section{Discussion and Future Work}
Despite the end-of-life (EOL) announcement in 2015, (\SI{33.81}{\percent}) of projects using Log4j continue to use Log4j 1.x, with only (\SI{10.89}{\percent}) migrating to Log4j-Core 2.x. This highlights persistent challenges in dependency management, including the reliance on outdated and vulnerable libraries.
This trend suggests a lack of awareness regarding the security risks of using an outdated framework or a perception that migration is prohibitively costly or complex. Continued reliance on Log4j 1.x poses significant security vulnerabilities, potentially leading to critical flaws if timely action is not taken.
The next step involves conducting a qualitative study to investigate the reasons why some projects do not migrate from obsolete libraries, while others transition quickly.
The next step is a qualitative analysis of why developers, especially why these deprecated libraries are still popular with even newer client systems.



\section*{Acknowledgment}
This work has been supported by JSPS KAKENHI Nos. JP20H05706, JP23K28065, JP23K16862, JP24K14895 and JST BOOST Grant Number JPMJBS2423.

\bibliographystyle{IEEEtranS.bst}
\bibliography{reference}
\end{document}